\documentclass[fleqn,usenatbib]{mnras}

\usepackage{newtxtext,newtxmath}

\usepackage[T1]{fontenc}
\usepackage{ae,aecompl}

\usepackage{graphicx}	
\usepackage{amsmath}	
\usepackage{amssymb}	

\title[Merger rate density of BBHs formed in OCs]{Merger rate density of binary black holes formed in open clusters}

\author[J. Kumamoto et al.]{
Jun Kumamoto,$^{1}$ \thanks{E-mail:kumamoto@astron.s.u-tokyo.ac.jp}
Michiko S. Fujii,$^{1}$ 
and Ataru Tanikawa,$^{2,3}$
\\
$^{1}$ Department of Astronomy, Graduate School of Science, The University of Tokyo, 7-3-1 Hongo, Bunkyo-ku, Tokyo 113-0033, Japan\\
$^{2}$ Department of Earth Science and Astronomy, College of Arts and Sciences, The University of Tokyo, 3-8-1 Komaba, Meguro-ku, Tokyo 153-8902, Japan \\
$^{3}$ RIKEN Center for Computational Science, 7-1-26 Minatojima-minami-machi, Chuo-ku, Kobe, Hyogo 650-0047, Japan
}

\date{Accepted XXX. Received YYY; in original form ZZZ}

\pubyear{2019}

\begin{document}
\label{firstpage}
\pagerange{\pageref{firstpage}--\pageref{lastpage}}
\maketitle

\begin{abstract}
Several binary black holes (BBHs) have been observed using gravitational wave detectors. For the formation mechanism of BBHs, two main mechanisms, isolated binary evolution and dynamical formation in dense star clusters, have been suggested. Future observations are expected to provide more information about BBH distributions, and it will help us to distinguish the two formation mechanisms. For the star cluster channel, globular clusters have mainly been investigated. However, recent simulations have suggested that BBH formation in open clusters is not negligible. We estimate a local merger rate density of BBHs originated from open clusters using the results of our $N$-body simulations of open clusters with four different metallicities. We find that the merger rate per cluster is the highest for our 0.1 solar metallicity model. Assuming a cosmic star formation history and a metallicity evolution with dispersion, we estimate the local merger rate density of BBHs originated from open clusters to be $\sim 70~{\rm yr}^{-1} {\rm Gpc}^{-3}$. This value is comparable to the merger rate density expected from the first and second observation runs of LIGO and Virgo. In addition, we find that BBH mergers obtained from our simulations can reproduce the distribution of primary mass and mass ratio of merging BBHs estimated from the LIGO and Virgo observations.
\end{abstract}



\section{Introduction}
\label{sec:intro}

The first two runs of gravitational wave detectors, LIGO and Virgo, have detected ten binary black hole (BBH) mergers \citep{2019PhRvX...9c1040A}. Most of the detected BBHs have masses of a few times $10M_{\odot}$ \citep{2016PhRvL.116f1102A, 2016PhRvL.116x1103A, 2017PhRvL.118v1101A, 2017PhRvL.119n1101A, 2017ApJ...851L..35A}. Before the detection of gravitational waves, black holes with such a mass was not known. The formation mechanism of these BBHs is still unclear.

There are two major scenarios for the origin of such BBHs. One is a common envelope and mass transfer evolution of isolated field binaries \citep[e.g.][]{1973NInfo..27....3T, 1998ApJ...506..780B, 2012ApJ...759...52D, 2014MNRAS.442.2963K, 2016Natur.534..512B, 2018MNRAS.474.2959G, 2019arXiv190612257B}. In this scenario, some heavy stars are born in binary. As the massive stars evolve, the orbital separation shrinks due to the common envelope. Therefore, BBHs formed via this process have a semi-major axis smaller than the initial binary separations.

The other scenario is the dynamical formation due to three-body encounters in the core of star clusters \citep{2000ApJ...528L..17P} or galactic nuclei \citep{2009MNRAS.395.2127O, 2016ApJ...831..187A}. The core of globular clusters ($10^5$--$10^6M_{\odot}$) has long been investigated as a formation site of BBHs in many previous works \citep[e.g.][]{2000ApJ...528L..17P, 2006ApJ...637..937O, 2008ApJ...676.1162S, 2010MNRAS.407.1946D, 2011MNRAS.416..133D, 2010MNRAS.402..371B, 2013MNRAS.435.1358T, 2014MNRAS.440.2714B, 2015PhRvL.115e1101R, 2016PhRvD..93h4029R, 2017PASJ...69...94F, 2017MNRAS.469.4665P, 2017MNRAS.464L..36A, 2018MNRAS.480.5645H, 2019ApJ...871...91Z}.

On the other hand, open cluster ($10^3$--$10^4M_{\odot}$) has not been expected as a main formation site of BBHs merging in the Hubble time because of the fewer number of massive stars and their shallower gravitational potential. However, the number of open clusters would have been an order of magnitude larger than that of globular clusters when they formed \citep{2010ARA&A..48..431P}. Although there may be few BBHs formed from one open cluster, more BBHs may be formed from open clusters throughout the Universe. Actually, the formation of BBHs in open clusters has been investigated by some previous works \citep[e.g.][]{2014MNRAS.441.3703Z, 2014ApJ...781...81G, 2016MNRAS.459.3432M, 2017MNRAS.467..524B, 2018MNRAS.473..909B, 2018MNRAS.481.5123B, 2019MNRAS.483.1233R, 2019MNRAS.487.2947D, 2019MNRAS.486.3942K, 2019ApJ...886...25B}.

In \citet[][hereafter Paper I]{2019MNRAS.486.3942K}, we found a new channel for the formation of merging BBHs in open clusters. In the case of the open cluster with the half-mass density of $10^4M_{\odot}~{\rm pc^{-3}}$, the core-collapse time is shorter than the lifetime of massive main-sequence. Then, massive main-sequence stars can form binaries in the dense core of the cluster before they evolve to black holes. Some of these binaries experience common envelope evolution and evolve to tight BBHs.

The merger rate density in the local Universe is calculated by integrating the merger rate density of BBHs ejected from clusters formed in each redshift \citep{2017PASJ...69...94F}. While most of the globular clusters are formed more than 12\,Gyr ago, open clusters are expected to form in any redshift. Therefore, the contribution of BBHs originated from the open cluster to the local merger rate density can be more significant. However, black hole mass strongly depends on metallicity, and we know that there is a cosmic metallicity evolution \citep{2017ApJ...840...39M}. Therefore, black holes formed in a lower redshift tend to be less massive due to stronger stellar wind of metal-richer star.

In order to investigate the local merger rate density of BBHs originated from open clusters in each redshift, we perform $N$-body simulations of open clusters with four different metallicity models. From the results of our simulations, we estimate local merger rate density of BBHs originated from open clusters formed at each cosmic time.

The structure of this paper is as follows. We describe our simulation methods and models in section 2. In Section 3, we investigate the properties of binary black holes formed in our simulations. In section 4, we calculate the merger rate from each model and estimate the local merger rate density from our simulation results. Conclusions are in sections 5.


\section{Methods and models}
\label{sec:methods}

We simulated open cluster models changing metallicity in addition to Model A in Paper I and investigated the formation of BBHs. A summary of our simulations is following.


\subsection{Initial conditions}

We set up four cluster models of which metallicities are $Z=0.002$, $0.005$, $0.01$ and $0.02$. Table \ref{tab:models} summarise our models. We set the initial cluster mass ($M_{\rm cl,ini}$) to be $2500~M_{\odot}$, which is the same as Model A in Paper I and named as Model Z0002 in the present paper. The number of runs ($N_{\rm run}$) per model depends on the metallicity. For more metal-rich models, we adopt a larger $N_{\rm run}$ because heavier BHs are expected to form less in more metal-rich clusters due to stronger stellar wind (see also subsection \ref{sec:StellarEvolution}).

\begin{table}
 \centering
 \caption{Models.}
 \label{tab:models}
 \begin{tabular}{lccc}
  \hline
               & $M_{\rm cl,ini} [M_\odot]$ & $Z$   & $N_{\rm run}$ \\
  \hline
  Model Z0002  & $2.5\times10^3$            & 0.002 & 360  \\
  Model Z0005  & $2.5\times10^3$            & 0.005 & 500  \\
  Model Z001   & $2.5\times10^3$            & 0.01  & 1000 \\
  Model Z002   & $2.5\times10^3$            & 0.02  & 1000 \\
  \hline
 \end{tabular}
\end{table}

As the initial density profile of cluster, we adopt Plummer profile \citep{1911MNRAS..71..460P};
\begin{equation}
    \rho(r) = \frac{3M_{\rm cl,ini}}{4\pi r_p^3} \left( 1+\frac{r^2}{r_p^2} \right)^{-5/2},
	\label{eq:plummer}
\end{equation}
\begin{equation}
    r_p = (2^{2/3}-1)^{1/2} r_{\rm hm},
	\label{eq:r_p}
\end{equation}
where $r_{\rm hm}$ is a half-mass radius. We set $r_{\rm hm}$ to be 0.31\,pc so that the initial half-mass density ($\rho_{\rm hm} = 3M_{\rm cl,ini}/8\pi r_{\rm hm}^3$) is $10^4 M_{\odot}~{\rm pc}^{-3}$. This half-mass density is similar to those of observed densest young massive clusters and higher than those of currently observed open clusters \citep{2010ARA&A..48..431P}. However, even if we set such an initial density higher than current values, open clusters experience core-collapse in a time shorter than the first supernova explosion (3--4\,Myr), and their densities immediately drop to the current density of observed typical open clusters \citep{2016ApJ...817....4F}. The half-mass radius is similarly increased from the initial value to the size of typical open clusters.

The initial mass of each stellar particle is given randomly from the Kroupa initial mass function \citep{Kroupa2001}. The lower and upper limit of the stellar mass are set to be $m_{\rm min} = 0.08 M_{\odot}$ and $m_{\rm max} = 150M_{\odot}$, respectively. In this case, the expected average stellar mass is $\langle m \rangle = 0.586M_{\odot}$. Thus, the initial number of particles is given as
\begin{equation}
    N_{\rm ini} = \frac{M_{\rm cl,ini}}{\langle m \rangle} = 4266.
	\label{eq:Nini}
\end{equation}

The half-mass relaxation time is calculated from the half-mass density as:
\begin{equation}
    t_{\rm rh} \sim 0.711\frac{N}{\log(0.4N)}\left(\frac{\rho_{\rm hm}}{M_{\odot}{\rm pc^{-3}}}\right)^{-0.5}~{\rm Myr}.
	\label{eq:t_rh}    
\end{equation}
The core-collapse time is correlated with the relaxation time, and the correlation factor depends on the ratio between the maximum and average masses of stars in the system. In our models, $m_{\rm max} / \langle m \rangle > 50$, and we obtain
\begin{equation}
  t_{\rm cc} \sim 0.07 t_{\rm rh,ini},
\end{equation}
from \citet{2004ApJ...604..632G,2016ApJ...817....4F}. Therefore, the core-collapse time of our models is estimated to be $\sim$0.7\,Myr.

We did not assume any primordial binaries. In Paper I, we discussed the effect of primordial binaries by comparing some previous studies for star clusters with similar mass range \citep{2014MNRAS.441.3703Z, 2019MNRAS.483.1233R, 2019MNRAS.487.2947D}. As a result, we argued that primordial binaries would not affect much on the formation rate of merging BBHs in open clusters for the following reason. The number of massive BHs ($\sim 20$--$30 M_{\odot}$) in open clusters is limited because of their total mass, and therefore, only a few massive BBHs can merge within the Hubble time. Even without primordial binaries, massive stars tend to form a massive binary in the core of star clusters. More discussion on this point is described in section 4.1 of Paper I.
Thus, we did not include primordial binaries in our simulation.


\subsection{$N$-body simulations}

We perform simulations of star clusters using a direct $N$-body simulation code, {\tt NBODY6++GPU} \citep{Wang+2015}. This code is an MPI-parallelised and GPU enabled version of {\tt NBODY6} \citep{1999PASP..111.1333A}. We perform our simulations using GPU cluster SGI Rackable C1102-GP8 (Reedbush-L) in the Information Technology Center, The University of Tokyo.

The motions of individual stars are integrated by a fourth-order Hermite scheme \citep{1992PASJ...44..141M}. In our simulations, binaries dynamically formed via three-body encounters. Some of these binaries have time steps much smaller than the time-scale of cluster evolution. Such hard binaries are integrated using KS regularization \citep{KustaanheimoStiefel1965, 1993CeMDA..57..439M}.

Since the core-collapse time of our models is $\sim$0.7\,Myr, most BBHs are formed within a few hundred Myr. In such a short time, tidal disruption of star clusters due to galactic tidal fields would not much affect the internal structure of star clusters. We, therefore, do not assume external tidal force.


\subsection{Stellar evolution}
\label{sec:StellarEvolution}

{\tt NBODY6++GPU} contains a stellar evolution model, {\tt SSE} \citep{Hurley+2000}. This model provides the time evolution of stellar radius, mass, and luminosity of each star depending on metallicity. We transported an updated mass loss model \citep{Belczynski+2010}, which is contained in the latest version of {\tt NBODY6}, to the stellar evolution model in {\tt NBODY6++GPU}. For binaries evolution, {\tt NBODY6++GPU} contains a binary evolution model \citep{1997MNRAS.291..732T}, which is an algorithm for rapid evolution binary star following the common envelope and mass transfer. We set common envelope efficiency parameter, $\alpha$, to be $1/3$.

\begin{figure}
	\includegraphics[width=\columnwidth]{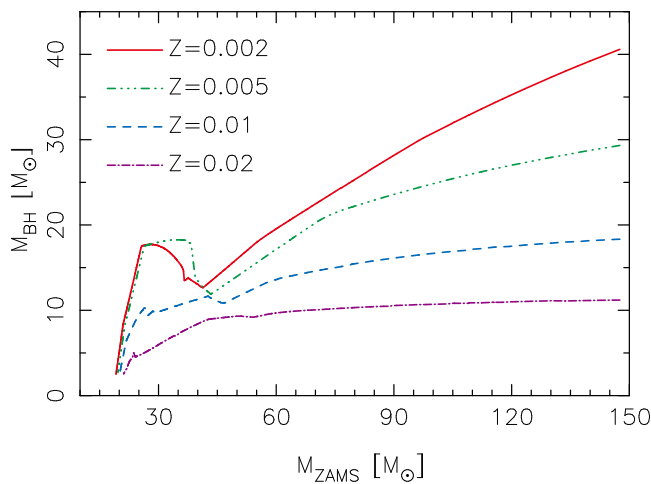}
    \caption{Relation between $M_{\rm ZAMS}$ and $M_{\rm BH}$ in the case of $Z=0.002$, $0.005$, $0.01$ and $0.02$.}
    \label{fig:MZAMS-MBH}
\end{figure}

The most important effect of metallicity is mass loss due to stellar winds. Therefore, the black hole mass also strongly depends on metallicity. Figure \ref{fig:MZAMS-MBH} shows relations between zero-age main-sequence stellar and black hole masses for each metallicity in our model. More metal-rich stars evolve into less massive black holes because of the stronger stellar wind.

In our simulation, we do not assume natal kicks caused by asymmetric supernovae explosion for simplification. The natal kicks may affect to the ejection rate of BBHs \citep{2013MNRAS.435.1358T}, but relatively massive black holes (10--$20~M_{\odot}$) tend to retain in star clusters \citep{2013ApJ...763L..15M}. Even if some black holes are ejected from the star clusters due to the natal kicks, the hardening of BBHs in open clusters mainly proceed via  interactions with the other stars rather than black holes. More discussion on this point is described in section 4.2 of Paper I.


\section{Properties of binary black holes}

We obtained in total $\sim300$--1000 BBHs ejected from cluster per model\footnote{We obtained not only BBHs but also black hole--main sequence binaries, which are one of the targets of the {\it Gaia} mission \citep{GaiaColl2016}. \cite{2020arXiv200111199S} investigated the detectability of these binaries by {\it Gaia} from our simulation results.}. In our analysis, we only analyzed the BBHs that ejected from clusters. There are also BBHs that remain in the cluster, but these have long semi-major axis and merger times longer than Hubble time. Therefore, ignoring these BBHs does not affect the our estimation of merger rate density of BBHs. In Table \ref{tab:BBHs}, we show the number of ejected BBHs ($N_{\rm BBH}$) formed in each model. About one BBH per one cluster formed in our simulations. We investigate the properties of these BBHs.

\begin{table}
 \centering
 \caption{Number of BBHs.}
 \label{tab:BBHs}
 \begin{tabular}{lcccc}
  \hline
              & $N_{\rm BBH}^{a}$ & $N_{\rm BBH}/N_{\rm run}$ & $N_{\rm mBBH}^{b}$ & $N_{\rm mBBH}/N_{\rm run}$ \\
  \hline
  Model Z0002 & 338 & 0.939 & 37 & 0.103 \\
  Model Z0005 & 487 & 0.974 & 17 & 0.034 \\
  Model Z001  & 988 & 0.988 & 32 & 0.032 \\
  Model Z002  & 877 & 0.877 &  7 & 0.007 \\
  \hline
  \multicolumn{5}{l}{\footnotesize$^a$ number of BBHs}\\
  \multicolumn{5}{l}{\footnotesize$^b$ number of merging BBHs}\\
 \end{tabular}
\end{table}

\begin{figure*}
	\includegraphics[width=0.96\textwidth]{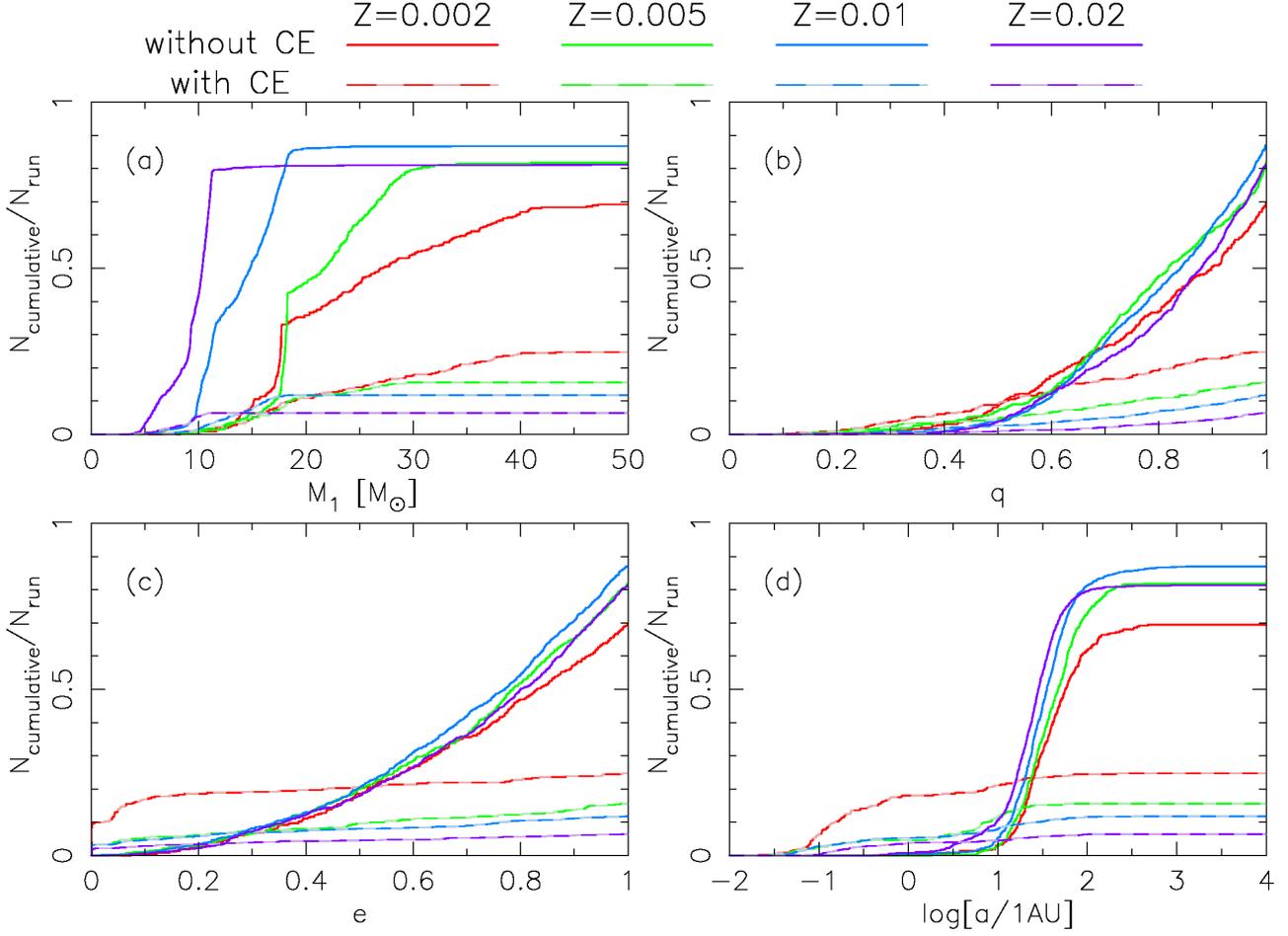}
    \caption{Cumulative distribution for $M_1$ (pnael (a)), $q$ (pnael (b)), $e$ (pnael (c)) and $a$ (pnael (d)) of BBHs formed in each metallicity model. Solid and dashed lines show the distribution of BBHs evolved with and without a common envelope.}
    \label{fig:Propertis_BBHs}
\end{figure*}

Figure \ref{fig:Propertis_BBHs} shows the cumulative distributions of BBHs formed (and ejected) in our simulations as a function of primary mass ($M_1$), mass ratio ($q$), orbital eccentricity ($e$), and semi-major axis ($a$). While dashed curves are for BBHs which experienced common envelope (including BBHs which experience the dynamical interaction after the common envelope.), solid ones are for BBHs which did not.

The cumulative distribution of the primary mass of ejected BBHs in panel (a) of Figure \ref{fig:Propertis_BBHs}. Primary mass of BBHs formed in metal-poorer model tends to be greater than that in the metal-richer model. This trend results from the black hole mass formed from massive main-sequence (See Figure \ref{fig:MZAMS-MBH}). The cumulative distributions of $q$ and $e$ do not show remarkable differences between metallicity models. Most of BBHs which experience common envelope has zero-eccentricity. These BBHs with non-zero-eccentricity are experienced the dynamical interaction after the common envelope.

Semi-major axis of BBHs which experience the common envelope tends to be shorter than those of BBHs without common envelope evolution. Interestingly, the semi-major axis of BBHs without common envelope formed in the metal-richer cluster model are shorter than those in metal-poorer cluster model. These binaries (binaries without common envelope evolution) formed via purely dynamical interactions. If the cluster mass is the same, the binding energy of ejected binaries should be similar among these models. Therefore, BBHs with a smaller $M_1$ tend to have a smaller value of $a$. Since the BH masses tend to be lower in more metal-rich clusters,  binaries dynamically formed in clusters with a higher-metallicity tend to have a shorter semi-major axis. 

We calculate merger time from the parameters of ejected binaries using the following equation \citep{1963PhRv..131..435P}:
\begin{eqnarray}
    t_{\rm GW} &=& \frac{5}{256} \frac{c^5}{G^3} \frac{a^4}{M_1^3q(1+q)} g(e) \\
    &\sim& 1.2 \left( \frac{M_1}{30M_{\odot}} \right)^{-3} \left( \frac{a}{0.1~{\rm AU}} \right)^{4} \frac{g(e)}{q(1+q)}~{\rm Gyr},
	\label{eq:tgw}
\end{eqnarray}
where
\begin{equation}
    g(e) = \frac{(1-e^2)^{3.5}}{1+(73/24)e^2+(37/96)e^4}.
\end{equation}
Here, $c$ and $G$ are light speed and gravitational constant, respectively.

\begin{figure*}
    \includegraphics[width=0.96\textwidth]{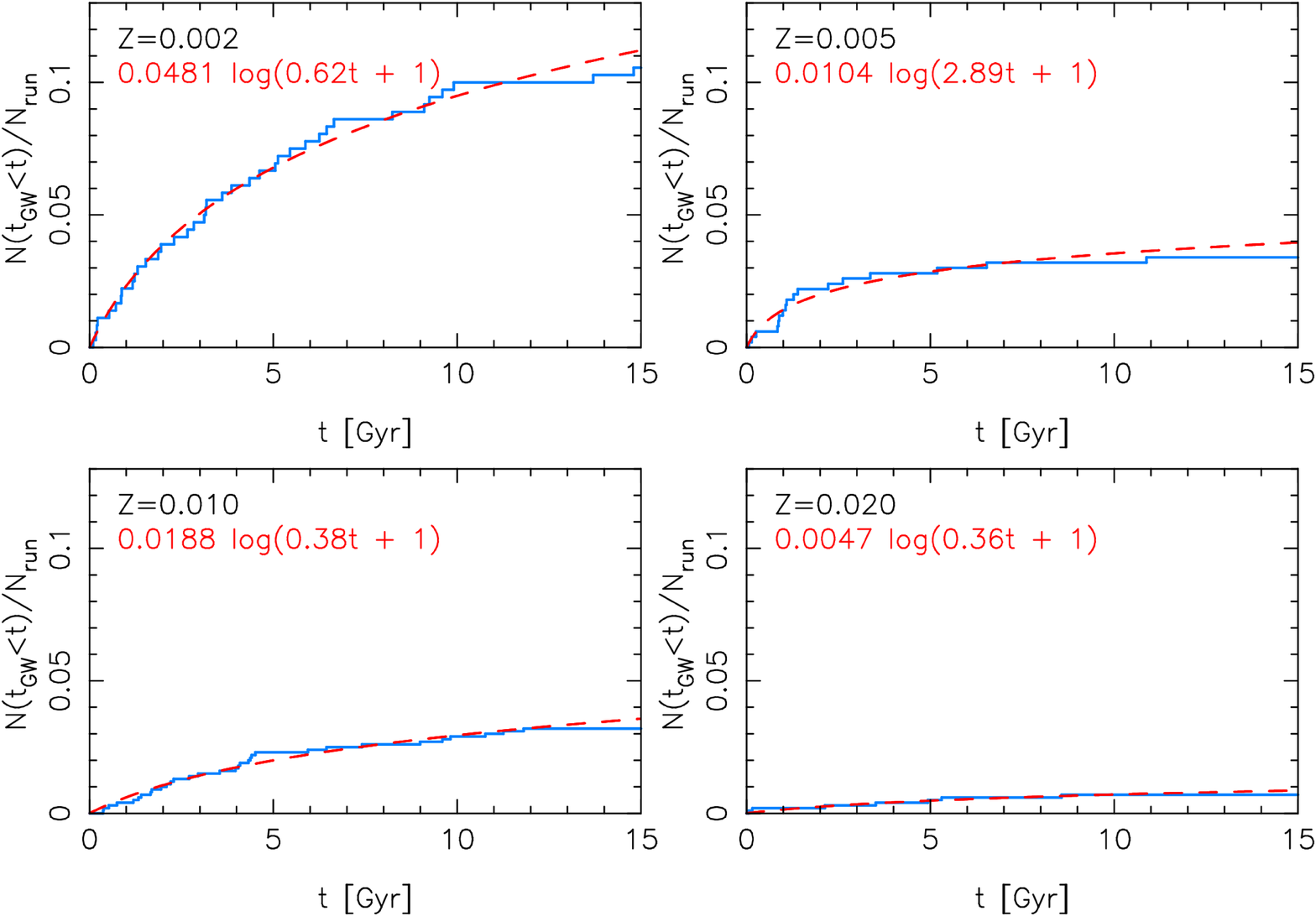}
    \caption{Cumulative distribution for the merger time of BBHs formed in each metallicity model. Dashed lines are fitting results, and its equations are shown in each panel. $t$ in the equation is normalised by Gyr.}
    \label{fig:CumulativeTgw}
\end{figure*}

Figure \ref{fig:CumulativeTgw} shows the cumulative distribution of the merger time for all models. Solid curves show the results from each model in our simulation. Dashed curves are fitted results using the following function: 
\begin{equation}
  N(t_{\rm GW}<t) = N_0 \ln\left(\tau^{-1} t+1\right),
  \label{eq:Nmerge}
\end{equation}
where $N_0$ and $\tau$ are fitting parameters. The fitted equations are also shown in each panel of Figure \ref{fig:CumulativeTgw}.

The number of BBHs merged within 14\,Gyr (hereafter, merging BBHs), $N_{\rm mBBH}$, in each model is summarised in Table \ref{tab:BBHs}. The number of merging BBHs per cluster is more significant in the metal-poorer model. 

\begin{figure*}
	\includegraphics[width=0.96\textwidth]{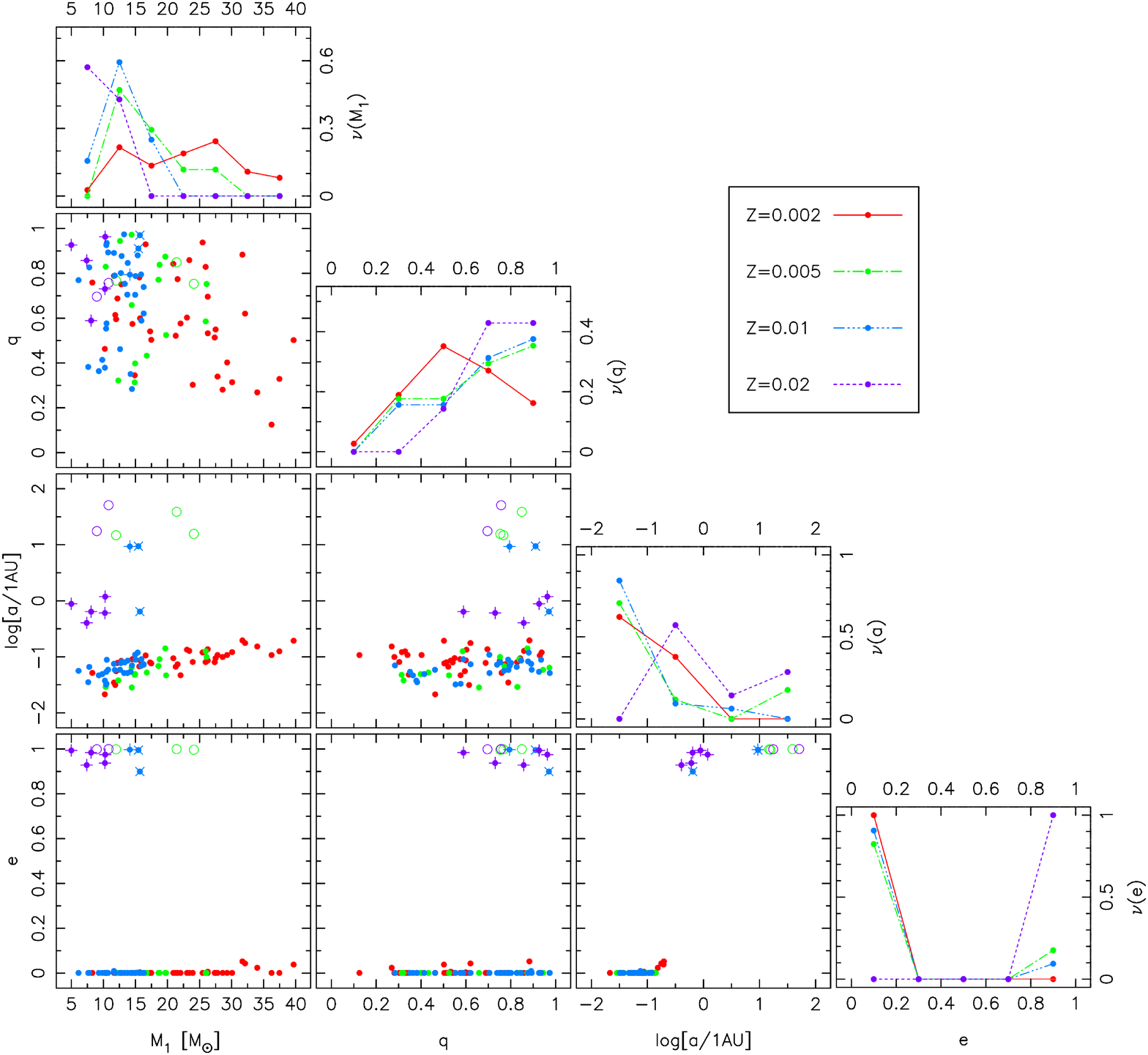}
    \caption{Distribution of merging BBHs. The six square panels show the relationship between $M_1$, $q$, $a$ and $e$ for the BBHs merging within 14\,Gyr. Filled and open circles show individual merging BBHs experienced and not experienced common envelope phase, respectively. Filled circles with ``$\times$'' indicate merging BBHs dynamically interacted with other stars after their common envelope phase. Filled circles with ``$+$'' indicate merging BBHs which experienced common envelope phase and then exchanged the member during interactions with a single black hole. The four rectangular panels drawn with dots and lines show the number fraction of merging BBHs calculated by equation (\ref{eq:nu}). The values are binned in each parameter. The bin size is consistent with the interval between tick marks on the horizontal axis for each panel.}
    \label{fig:mergingBBH}
\end{figure*}

Figure \ref{fig:mergingBBH} shows the primary mass ($M_1$), mass ratio ($q$), semi-major axis ($a$), and eccentricity ($e$) of merging BBHs ejected from open clusters. BBHs which experienced common envelope have a distribution different from those which did not. In general, merging BBHs experienced common envelope phase have a short semi-major axis and nearly zero eccentricity. However, some BBHs which experienced common envelop phase dynamically interact with other single stars, and as a result, the eccentricity is pumped up. These eccentric BBHs can merge within 14\,Gyr because of their high eccentricity (see eq. (\ref{eq:tgw})). Furthermore, some of them result in an exchange of a member with an encountering single black hole during three-body encounters. 

Merging BBHs which experienced dynamical interactions after their common envelope phase had a relatively large semi-major axis compared with those which can merge just after the common envelope evolution. Because of their relatively large cross-section, they could interact with other stars and change their orbital parameters. The semi-major axis of BBHs which experienced common envelope phase distribute $-1.5 \lesssim \log a \lesssim 2$ (panel (d) of Figure \ref{fig:Propertis_BBHs}). Among them, only BBHs which could experience dynamical interactions and get an eccentricity high enough to merge within 14\,Gyr appear as a merging BBH.

We also find a few merging BBHs which did not experience common envelope evolution (dynamically formed BBHs). All of these merging BBHs have a semi-major axis larger than those of merging BBHs which experienced common envelope phase. The eccentricities of the dynamically formed BBHs are almost one (actually, $e>0.995$ in our results). Because of such high eccentricities, they can merge within 14\,Gyr in spite of their relatively large semi-major axis.

In Figure \ref{fig:mergingBBH}, we also show the number distribution of merging BBHs for each parameter; 
\begin{equation}
    \nu=\frac{N_{\rm bin}}{N_{\rm mBBH}},
    \label{eq:nu}
\end{equation}
where $N_{\rm bin}$ is the number of merging BBH included each parameter bin. The bin sizes are equal to $5M_{\odot}$, 0.2, 1 and 0.2 for $M_1$, $q$, $\log(a)$ and $e$.


\section{Local merger rate density}
\subsection{Integration}
A local merger rate density of BBHs originated from star clusters is written as,
\begin{equation}
    R = \int dM_{\rm cl} \int dZ \int dt_{\rm L} D(Z, M_{\rm cl}, t_{\rm GW}=t_{\rm L}) \frac{d\dot{N}_{\rm cl}}{dZdM_{\rm cl}}(t_{\rm L}, Z, M_{\rm cl}).
    \label{eq:merger_rate}
\end{equation}
where $t_{\rm L}$ is lookback time, and $D(Z, M_{\rm cl}, t_{\rm GW})$ is the merger rate of BBHs originated from one cluster, which is having metallicity, $Z$, and mass, $M_{\rm cl}$, and merging after $t_{\rm GW}$ from the cluster formation. In order to obtain a current merger rate, we need to count merger rates of BBHs with $t_{\rm GW}=t_{\rm L}$. In equation (\ref{eq:merger_rate}), $d\dot{N}_{\rm cl}(t_{\rm L}, Z, M_{\rm cl})/dZdM_{\rm cl}$ is the formation rate density of cluster with a metallicity of $Z$ and a mass of $M_{\rm cl}$.

\begin{figure*}
    \includegraphics[width=0.96\textwidth]{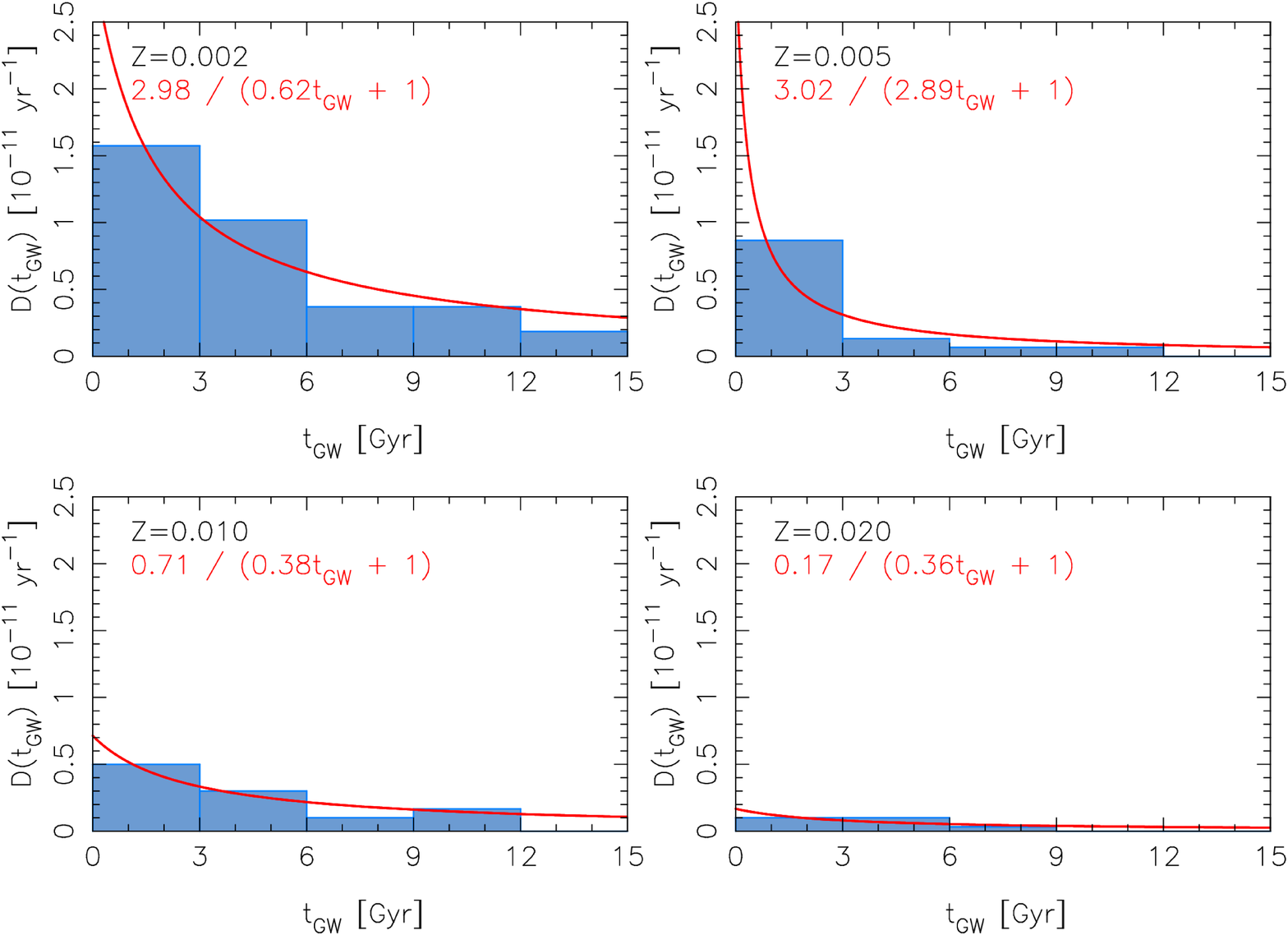}
    \caption{Delayed-time distribution, $D(Z, M_{\rm cl}, t_{\rm GW})$ in the case of $Z=0.002$, 0.005, 0.010 and 0.020, and $M_{\rm cl}=2500M_\odot$. Histograms are the results of our simulation. Solid lines are gotten by the differential of fitting results in figure \ref{fig:CumulativeTgw}, and its equations are shown in each panel. In these equations, $t_{\rm GW}$ is normalised by Gyr.}
    \label{fig:DTD}
\end{figure*}

We first derive $D(Z, M_{\rm cl}, t_{\rm GW})$ from the delayed-time distribution obtained from our simulations. From the fitted function to the cumulative distribution of merger time shown in Figure \ref{fig:CumulativeTgw}, we can obtain the delayed-time distribution for each metallicity model as
\begin{equation}
  D(Z, M_{\rm cl}=2500M_\odot, t_{\rm GW}) = \frac{\mathrm{d}N(t_{\rm GW}<t)}{\mathrm{d}t} \times \frac{1}{N_{\rm run}}.
\end{equation}
In Figure \ref{fig:DTD}, we show the equation for each metallicity and those obtained from the distribution of BBHs formed in our simulations. We confirmed that the fitted functions are consistent with the simulation results. 

Next, we estimate the formation rate density of clusters. The following equation shows the formation rate density of the cluster with a mass of $M_{\rm cl}$;
\begin{equation}
    \frac{d\dot{N}_{\rm cl}}{dZ dM_{\rm cl}}(t_{\rm L}, Z, M_{\rm cl}) = \frac{f_{\rm cl}(M_{\rm cl})}{M_{\rm cl}} \frac{d\Psi(t_{\rm L}, Z)}{dZ}.
    \label{eq:Psi}
\end{equation}
Here, $\Psi(t_{\rm L}, Z)$ is a comoving formation rate density of star which is having metallicity, $Z$, and $f_{\rm cl}(M_{\rm cl})$ is the fraction of stellar mass formed as star cluster having the mass, $M_{\rm cl}$. If we assume that all stars are formed as members of clusters with a mass range of $10^2$ to $10^6$, then we can write as 
\begin{equation}
    \int_{10^2M_{\odot}}^{10^6M_{\odot}} f_{\rm cl}(M_{\rm cl}) dM_{\rm cl} = 1.
\end{equation}
In addition, we assume that the cluster mass function follow to $M_{\rm cl}^-2$ ($f_{\rm cl}(M_{\rm cl})/M_{\rm cl} \propto M_{\rm cl}^{-2}$), therefore,
\begin{equation}
    f_{\rm cl}(M_{\rm cl}) = \frac{1}{4\ln{10}} M_{\rm cl}^{-1}.
\end{equation}
With this assumption, we estimate $R_{\rm OC}$, which is the local merger rate density of BBHs formed in open clusters with a mass of $10^3M_{\odot}$ to $10^4M_{\odot}$, as
\begin{equation}
    R_{\rm OC} = \frac{1}{4\ln{10}} \int_{10^3M_{\odot}}^{10^4M_{\odot}} dM_{\rm cl} \int dZ \int dt_{\rm L} \frac{D(Z, M_{\rm cl}, t_{\rm L})}{M_{\rm cl}^2} \frac{d\Psi(t_{\rm L}, Z)}{dZ}.
    \label{eq:R_oc0}
\end{equation}
Here, we assume that the dependence of $D(Z, M_{\rm cl}, t_{\rm L})$ on $M_{\rm cl}$ is negligible in this cluster mass range, we can rewrite equation (\ref{eq:R_oc0}) as
\begin{equation}
    R_{\rm OC} = \frac{9\times10^{-4}}{4M_{\odot}\ln{10}} \int dZ \int dt_{\rm L} D(Z, t_{\rm GW}=t_{\rm L}) \frac{d\Psi(t_{\rm L}, Z)}{dZ}.
    \label{eq:R_oc}
\end{equation}
Hereafter, we omit the argument $M_{\rm cl}$ in function $D(Z, M_{\rm cl}, t_{\rm GW})$ for simplification. In order to calculate the integral part of this equation, we apply two methods. The first one is to assume that all star clusters which were born in the same era have the same averaged metallicity at that time (single metallicity evolution model). The other is to consider metallicity dispersion in each formation era as well as the averaged metallicity evolution (metallicity dispersion model). The merger rate density calculated using each model will be described in subsection \ref{sec:noZdistribution} and \ref{sec:Zdistribution}, respectively. In addition, we calculate the differential merger rate density of $M_1$ and $q$ for each model.

\subsection{Single metallicity evolution model}
\label{sec:noZdistribution}

First, we assume that all stars born at the same era have the same metallicity and that the metallicity evolve with time. In order to connect the metallicity of star clusters to their birth time, we use a relation between metallicity ($Z$) and redshift ($z$) \citep{2017ApJ...840...39M},
\begin{equation}
    \log (Z/Z_{\odot}) = 0.153-0.074z^{1.34}.
    \label{eq:Z-z}
\end{equation}
Thus, stars formed at any lookback time ($t_{\rm L}$) have the metallicity ($Z(t_{\rm L})$), so that,
\begin{equation}
    \frac{d\Psi(t_{\rm L}, Z)}{dZ} = \Psi(t_{\rm L}) \delta(Z(t_{\rm L})),
\end{equation}
where $\delta$ is the Dirac delta function. Substituting this for equation (\ref{eq:R_oc}), we obtain
\begin{eqnarray}
    R_{\rm OC} &=& \frac{9\times10^{-4}}{4M_{\odot}\ln{10}} \int dZ' \int dt_{\rm L} D(Z', t_{\rm L}) \Psi(t_{\rm L}) \delta(Z(t_{\rm L})) \nonumber \\
        &=& \frac{9\times10^{-4}}{4M_{\odot}\ln{10}} \int dt_{\rm L} D(Z(t_{\rm L}), t_{\rm L}) \Psi(t_{\rm L}). 
    \label{eq:R_oc1}
\end{eqnarray}
The star formation rate (SFR) density, $\Psi(t_{\rm L})$, is given as a function of redshift ($z$) \citep{2017ApJ...840...39M}, such as: 
\begin{equation}
    \Psi(z) = 0.01 \frac{(1+z)^{2.6}}{1+((1+z)/3.2)^{6.2}} M_{\odot} {\rm yr}^{-1} {\rm Mpc}^{-3}.
    \label{eq:SFRdensity}
\end{equation}


\subsubsection{Merger rate density for each metallicity}
We calculate $D(Z,t_{\rm L}) \Psi(t_{\rm L})$ from our simulations and equation (\ref{eq:SFRdensity}). Although $D$ is a function of metallicity ($Z$), it is known that the typical metallicity in the Universe evolves. We can calculate the local merger rate of BBHs, originated from those clusters, per cluster, $D(Z, t_{\rm GW}=t_{\rm L})$, from the functions shown in Figure \ref{fig:DTD}. Table \ref{tab:results} shows the results of these calculations. We obtain $3.3\times10^{-12}~{\rm yr}^{-1}$ for Model Z0002, and this value is the largest among our models. The value of $D(Z,t_{\rm L})$ tends to decrease as metalicity increase. For Model Z002, $D(0.02,t_{\rm L})$ is only 10\% of that for Model Z0002.

\begin{table*}
 \centering
 \caption{Our results ($M_{\rm cl} = 2500M_{\odot}$).}
 \label{tab:results}
 \begin{tabular}{lcccccc}
  \hline
              & $Z$ & $z^a$ & $t_{\rm L}^b$ & $D(Z,M_{\rm cl}=2500M_\odot,t_{\rm GW}=t_{\rm L})^c$ & $ \Psi(z)^d $  & $D(Z,2500M_\odot,t_{\rm L})\Psi(z)$\\
              & & & [Gyr] & $[10^{-11} {\rm yr}^{-1}]$ & $[M_{\odot} {\rm yr}^{-1} {\rm Mpc}^{-3}]$ & $[10^{-5} M_{\odot} {\rm yr}^{-2} {\rm Gpc}^{-3}]$\\
  \hline
  Model Z0002 & 0.002 & 7.76 & 13.1 & 0.33  & 0.00547 & 1.8 \\
  Model Z0005 & 0.005 & 5.66 & 12.8 & 0.081 & 0.0145  & 1.2 \\
  Model Z001  & 0.01  & 3.87 & 12.2 & 0.13  & 0.0422  & 5.4 \\
  Model Z002  & 0.02  & 1.72 & 9.97 & 0.037 & 0.0988  & 3.7 \\
  \hline
  \multicolumn{7}{l}{\footnotesize$^a$ redshift calculated from $Z$ and equation~(\ref{eq:Z-z})}\\
  \multicolumn{7}{l}{\footnotesize$^b$ lookback time at redshift $z$ in the case of $(h,\Omega_M, \Omega_\Lambda)=(0.7,0.3,0.7)$}\\
  \multicolumn{7}{l}{\footnotesize$^c$ local merger rate calculated from functions shown in figure \ref{fig:DTD} for each metallicity}\\
  \multicolumn{7}{l}{\footnotesize$^d$ SFR density calculated from equation~(\ref{eq:SFRdensity})}
 \end{tabular}
\end{table*}

On the other hand, SFR density ($\Psi(z)$) has a peak at $z\simeq2.0$ ($t_{\rm L}\simeq 10.5$\,Gyr, see Figure \ref{fig:DPsi}). The value of $\Psi (z)$ when Model Z001 like cluster was born ($z=3.87$, $t_{\rm L}=12.2$\,Gyr), is about 20 times greater than that for Model Z0002 ($z=7.76$, $t_{\rm L}=13.1$\,Gyr). Therefore, the contribution to the local merger rate of the metal-richer cluster may be greater because of the more active star formation.

We calculate $D(Z,t_{\rm L}) \Psi(z)$ for each metallicity in our simulations, and the results are shown in Table \ref{tab:results}. These values present the contribution to the local merger rate from clusters which have metallicity $Z$ and are born in unit time and unit volume. If we compare the total merger rate including cosmic SFR density, the contribution from metal-rich clusters (Models Z001 and Z002) is larger than that from metal-poor clusters (Models Z0001 and Z0002) due to the cosmic SFR density.


\subsubsection{Local merger rate density}

\begin{figure}
	\includegraphics[width=\columnwidth]{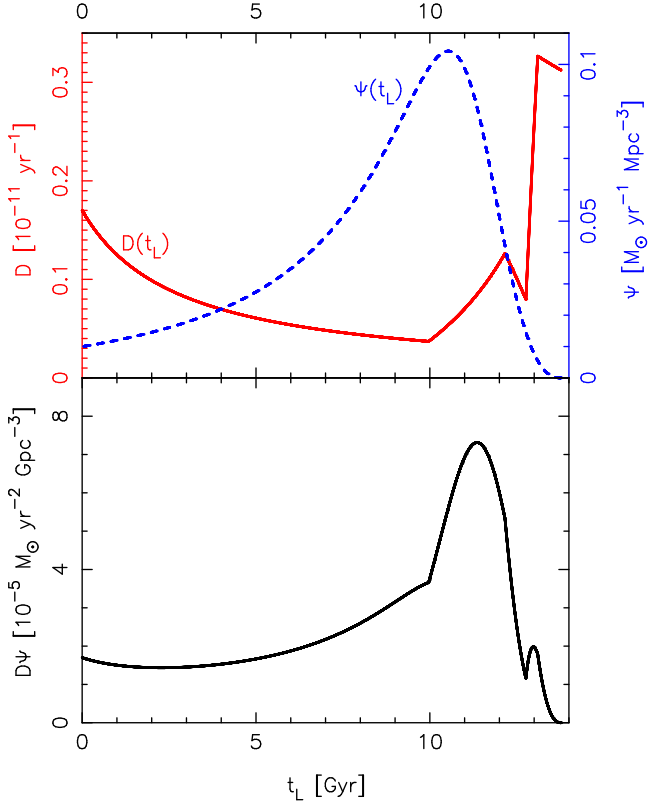}
	\caption{Top panel shows $D(Z,t_{\rm GW}=t_{\rm L})$ (solid line) and $\Psi(t_{\rm L})$ (dashed line) calculated from equation (\ref{eq:approximatedD}) and (\ref{eq:Psi}) as a function of $t_{\rm L}$. Bottom panel shows the product of $D(Z,t_{\rm L})$ and $\Psi(t_{\rm L})$.}
	\label{fig:DPsi}
\end{figure}

In order to calculate the equation (\ref{eq:R_oc1}), we have to integrate $D(Z(t_{\rm L}),t_{\rm L})\Psi(t_{\rm L})$. However, the delayed distribution function $D(Z(t_{\rm L}),t_{\rm L})$ is dispersively given from our simulations. We, therefore, interpolate and extrapolate the data points obtained from four our simulation models as follows:
\begin{equation}
    \begin{aligned}
        & D(Z, t_{\rm GW}) = \\
        & \left\{
        \begin{aligned}
            & D_1 &\text{for}& \quad Z < 0.002, \\
            & \frac{0.005-Z}{0.003}D_1+\frac{Z-0.002}{0.003}D_2 &\text{for}& \quad 0.002 \leqq Z < 0.005, \\
            & \frac{0.01-Z}{0.005}D_2+\frac{Z-0.005}{0.005}D_3 &\text{for}& \quad 0.005 \leqq Z < 0.01, \\
            & \frac{0.02-Z}{0.01}D_3+\frac{Z-0.01}{0.01}D_4 &\text{for}& \quad 0.01 \leqq Z < 0.02, \\
            & D_4 &\text{for}& \quad 0.02 \leqq Z.
        \end{aligned}
        \right.
    \end{aligned}
    \label{eq:approximatedD}
\end{equation}
Here,
\begin{eqnarray}
    D_1 &=& D(Z=0.002,t_{\rm GW}), \\
    D_2 &=& D(Z=0.005,t_{\rm GW}), \\
    D_3 &=& D(Z=0.01,t_{\rm GW}), \\
    D_4 &=& D(Z=0.02,t_{\rm GW}).
\end{eqnarray}
The interpolated and extrapolated $D(Z,t_{\rm GW}=t_{\rm L})$ as a function of $t_{\rm L}$ is shown by the solid curve in the top panel of Figure \ref{fig:DPsi}. In the same panel, the dashed line shows the SFR density, $\Psi(t_{\rm L})$. 

The bottom panel of Figure \ref{fig:DPsi} shows $D(Z,t_{\rm GW}=t_{\rm L})\Psi(z(t_{\rm L}))$ as a function of $t_{\rm L}$. This distribution has a maximum value between $t_{\rm L} = 11$ and 12\,Gyr and decreases toward the present time because the SFR density decreases. There is a small peak at $t_{\rm L} \sim 13~{\rm Gyr}$. This peak results from the distribution of $D(Z,t_{\rm L})$, in which $D(Z=0.01, t_{\rm L})$ is larger $D(Z=0.005, t_{\rm L})$ contrary to the number of merging BBH per cluster decreases as metallicity increases. This may be due to the relatively small number of runs. We obtained tens of merging BBHs for each model, but these may not be enough to fit a function to obtain delayed-time distribution (see Fig. \ref{fig:DTD}). However, the contribution to the estimation of the merger rate density would not be large even if the values of $D(Z, t_{\rm L})$ slightly changed.

We finally integrate $D(Z,t_{\rm L})\Psi(z(t_{\rm L}))$ and estimate the local merger rate density of BBHs ejected from open clusters using equation (\ref{eq:R_oc1}) as
\begin{equation}
    R_{\rm OC} \sim 35~{\rm yr^{-1}Gpc^{-3}}.
\end{equation}
We apply the delayed distribution function at $Z=0.02$ to calculate merger rate density for $t_{\rm L} < 9.97~{\rm Gyr}$, although the averaged metallicity of this time is larger than $Z=0.02$. Because of this, we might overestimate $R_{\rm OC}$. On the other hand, \cite{2019ApJ...882L..24A} have estimated the BBH merger rate density from the first and second observing runs of LIGO and Virgo as
\begin{equation}
    R_{\rm obs}=64.0_{-33.0}^{+73.5}~{\rm yr}^{-1}{\rm Gpc}^{-3},
    \label{eq:Robs}
\end{equation}
in their Model A. Our estimation corresponds to $\sim$55\,\% of this value.


\subsubsection{Differential merger rate density}

\begin{figure}
	\includegraphics[width=\columnwidth]{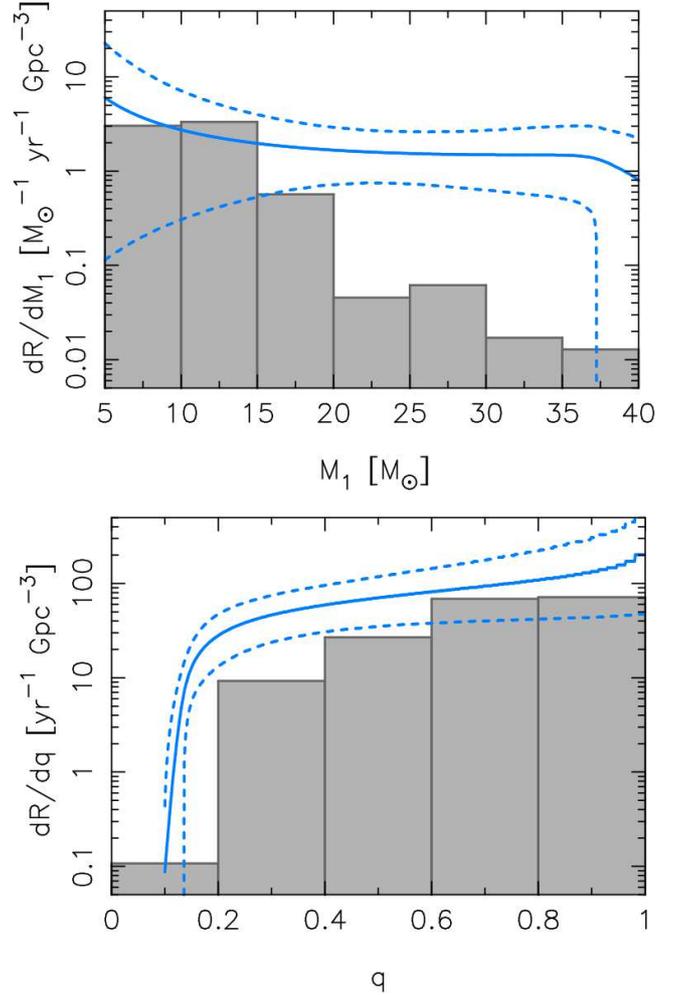}
	\caption{Differential merger rate density for $M_1$ (top panel) and $q$ (bottom panel). Histograms show our results calculated using equations (\ref{eq:dRdM1}) and (\ref{eq:dRdq}). The solid curve shows the posterior probability distribution of differential merger rate density, expected from 10 BBH mergers in O1 and O2 of LIGO and Virgo \citep[Model A of][]{2019ApJ...882L..24A}. The region between the two dashed lines shows the 90\% confidence intervals.}
	\label{fig:dR}
\end{figure}

We also give the distribution of $M_1$ and $q$ of BBHs expected to merge in the local Universe. From equations (\ref{eq:R_oc}) and (\ref{eq:nu}), the differential merger rate density is obtained as
\begin{equation}
    \frac{dR_{\rm OC}}{dM_1} = \frac{1}{W_{M_1}} \frac{9\times10^{-4}}{4M_{\odot}\ln{10}} \int dt_{\rm L} \nu (M_1) D(Z(t_{\rm L}), t_{\rm L}) \Psi(t_{\rm L}),
    \label{eq:dRdM1}
\end{equation}
and
\begin{equation}
    \frac{dR_{\rm OC}}{dq} = \frac{1}{W_{q}} \frac{9\times10^{-4}}{4M_{\odot}\ln{10}} \int dt_{\rm L} \nu (q) D(Z(t_{\rm L}), t_{\rm L}) \Psi(t_{\rm L}),
    \label{eq:dRdq}
\end{equation}
for $M_1$ and $q$, respectively. Here, $W_{M_1}$ and $W_q$ are the bin sizes for $\nu(M_1)$ and $\nu(q)$, respectively. The results obtained from our simulations are shown as histograms in Figure \ref{fig:dR}. We also plot the differential merger rate density expected from 10 BBH mergers detected in O1 and O2 of LIGO and Virgo \citep[Model A of][]{2019ApJ...882L..24A}.

For $M_1$, our result shows a good agreement with the distribution estimated from the observed BBH mergers at the low mass end. However, the merger rate of massive ($M_1\gtrsim20M_\odot$) BBHs predicted from our simulation is much smaller than those expected from the observations. For $q$, our differential merger rate densities are less than those of observation in most of $q$, but the shapes are similar.

Here, the metallicity of star clusters was adopted by the mass-weighted average metallicity at each $t_{\rm L}$. However, the metallicity dispersion at each $t_{\rm L}$ should also be considered. In lower-metallicity clusters, the merger rate of BBHs from a cluster is larger than that of higher-metallicity clusters. In addition, the distribution of $M_1$ is flatter, if the metallicity is lower. Therefore, the models with metallicity dispersion may increase the total merger rate density and the number of merging BBHs with a larger $M_1$.


\subsection{Metallicity dispersion model}
\label{sec:Zdistribution}

\subsubsection{Local merger rate density with metallicity dispersion}

\begin{figure}
	\includegraphics[width=\columnwidth]{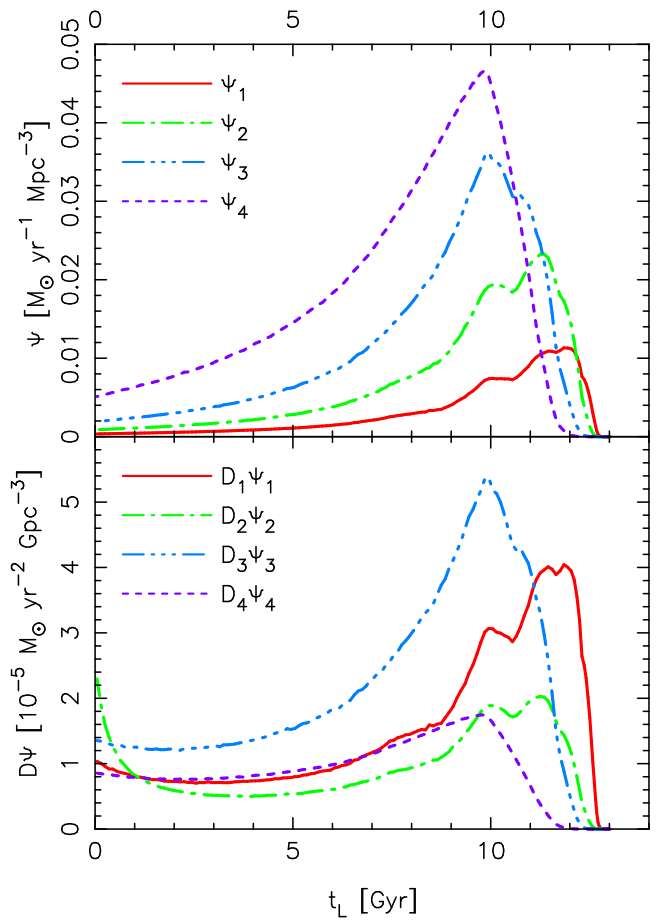}
	\caption{Top panel shows the SFR density for stars having each metallicity as a function of the lookback time. Bottom panel shows the product of $D_i(t_{\rm L})$ and $\Psi_i(t_{\rm L})$}.
	\label{fig:DPsi2}
\end{figure}

Next, we estimate a local merger rate density with a cosmic metallicity evolution and the dispersion at each time. We add the metallicity dispersion by considering cosmic star formation density for stars with each metallicity and then integrating the results with respect to metallicity. The result should give us the total merger rate density with metallicity dispersion at each cluster formation time. With this assumption, $R_{\rm OC}$ in equation (\ref{eq:R_oc}) is approximated as; 
\begin{equation}
    R_{\rm OC} \sim \frac{9\times10^{-4}}{4M_{\odot}\ln{10}} \sum_{i=1}^4 \int dt_{\rm L} D_i(t_{\rm GW}=t_{\rm L}) \Psi_i(t_{\rm L}),
    \label{eq:R_oc2}
\end{equation}
where
\begin{eqnarray}
    \Psi_1(t_{\rm L}) = \int_{0.001}^{0.002} dZ \frac{d\Psi}{dZ}(Z, t_{\rm L}), \\
    \Psi_2(t_{\rm L}) = \int_{0.002}^{0.005} dZ \frac{d\Psi}{dZ}(Z, t_{\rm L}), \\
    \Psi_3(t_{\rm L}) = \int_{0.005}^{0.01} dZ \frac{d\Psi}{dZ}(Z, t_{\rm L}), \\
    \Psi_4(t_{\rm L}) = \int_{0.01}^{0.02} dZ \frac{d\Psi}{dZ}(Z, t_{\rm L}).
\end{eqnarray}
Here, we divided the metallicity to four regions, for which we performed $N$-body simulations. By integrating these functions with respect to $Z$, we can obtain the SFR density evolution $\Psi_i(t_{\rm L})$ for each metallicity. In order to calculate $\Psi_i$, we use publicly available data of $d\Psi(Z, t_{\rm L})/dZ$ shown in Figure 6 of \citet{2019MNRAS.488.5300C} (\url{https://ftp.science.ru.nl/astro/mchruslinska/}). The obtained $\Psi_1(t_{\rm L})$--$\Psi_4(t_{\rm L})$ are shown in the top panel of Figure \ref{fig:DPsi2}. Using $\Psi_i(t_{\rm L})$ and $D_i(t_{\rm L})$ obtained from our simulations, we can calculate $D_i(t_{\rm L}) \Psi_i (t_{\rm L})$, which are shown in the bottom panel of Figure \ref{fig:DPsi2}. 

The SFR density with $0.01<Z<0.02$ (the highest metallicity range), i.e., $\Psi_4(t_{\rm L})$, is larger than any other $\Psi_i(t_{\rm L})$ at $t_{\rm L}<10.5$\,Gyr (see the top panel of Figure \ref{fig:DPsi2}). However, the value of $D(t_{\rm L})Psi(t_{\rm L})$ is the largest for the metallicity of $0.005<Z<0.01$, i.e., $D_3(t_{\rm L})\Psi_3(t_{\rm L})$, at $0.4~{\rm Gyr}<t_{\rm L}<11.1~{\rm Gyr}$ (see the bottom panel of Figure \ref{fig:DPsi2}). In addition, $D_4(t_{\rm L})\Psi_4(t_{\rm L})$ has a peak at $t_{\rm L}>11.1~{\rm Gyr}$, but in later time ($t_{\rm L}<9~{\rm Gyr}$), it is comparable to the value of $D_1(t_{\rm L})\Psi_1(t_{\rm L})$. Unlike the case without metallicity dispersion (see subsection \ref{sec:noZdistribution}), the contribution from low-metallicity clusters born in a low redshift is significant, because a non-negligible fraction of star clusters with low-metallicity continues to be formed even in a low redshift.

We finally estimate $R_{\rm OC}$ from equation (\ref{eq:R_oc2}) and $D_i(t_{\rm L})\Psi_i(t_{\rm L})$ shown in Figure \ref{fig:DPsi2} as
\begin{equation}
    R_{\rm OC} \sim 70~{\rm yr^{-1}Gpc^{-3}}.
\end{equation}
This value is twice as large as that obtained in the case without metallicity dispersion and in good agreement with that estimated from the observations (equation (\ref{eq:Robs})). Note that this result is estimate extrapolated from our star cluster models ($M_{\rm ini}=2500 M_{\odot}$ and $\rho_{\rm hm,ini}=10^4 M_{\odot}{\rm pc}^{-3}$). In order to estimate a more realistic value, we should consider the spectrum of initial mass and size of star clusters.


\subsubsection{Differential merger rate density with metallicity dispersion}

\begin{figure}
	\includegraphics[width=\columnwidth]{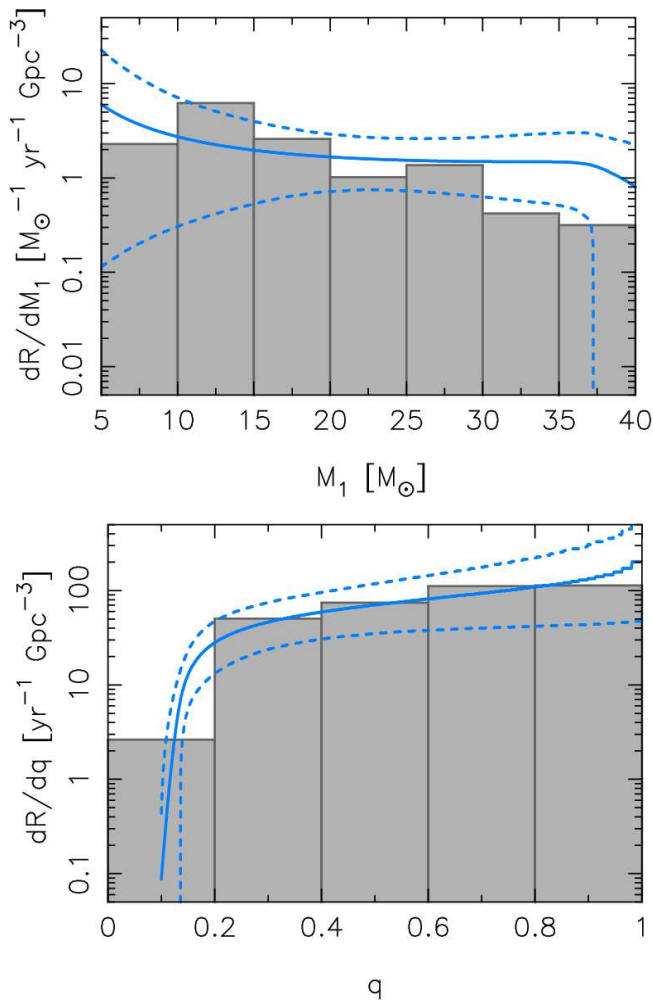}
	\caption{Same as Figure \ref{fig:dR}, but for the model considering metallicity dispersion.}
	\label{fig:dR2}
\end{figure}

We calculate the differential merger rate density for $M_1$ and $q$ in the case with metallicity dispersion similar to those in subsection \ref{sec:noZdistribution}. The results are shown in Figure \ref{fig:dR2}.

The top panel shows $dR_{\rm OC}/dM_1$. Compared to the case without metallicity dispersion (see Figure \ref{fig:dR}), we find a larger number of merging BBHs for larger $M_1$. This is caused by the BBHs originated from lower-metallicity clusters formed at low-$z$, which were not considered the model without metallicity dispersion. Between $20M_{\odot}$ and $30M_{\odot}$, our results are consistent with the observation. Above $30M_{\odot}$, however, the differential merger rate that we obtained is still several times smaller than that estimated from observed BBH mergers. The lack of massive BBHs in our models can be filled by BBHs originated from globular clusters \citep{2016PhRvD..93h4029R,2017PASJ...69...94F}, and/or isolated binaries\citep{2015ApJ...806..263D}, formed in high redshift, at which the expected BH mass is higher.

Considering the metallicity dispersion, the differential merger rate density of $q$ increased overall and is consistent with the estimation from the observations, but still less at $q\sim 1$. BBHs with $q\sim 1$ tend to be formed in globular clusters and the field (isolated binary evolution), which we did not take into account in this study. Even if the number of equal-mass merging BBHs increases a few times when we add BBH populations formed in globular clusters and the field, the merger rate density would be consistent within the 90\% confidence interval of the observation.


\section{Conclusions}

We performed a series of $N$-body simulation of open clusters with four different metallicity models, $Z=0.002$, 0.005, 0.01, and 0.02. Roughly one BBH per one cluster was formed in all cluster models. 

We investigated the properties of BBHs formed in our simulations. In our simulations, the primary mass of BBHs formed in metal-richer open clusters tends to be smaller than that formed in metal-poorer clusters because of stronger stellar-wind mass-loss of metal-richer stars. The number of BBHs merging within cosmic time is the highest for the $Z = 0.002$ model. We also investigated the delayed-time distribution function of the merger time of BBHs formed in open cluster with each metallicity.

We also estimated a local merger rate density, considering the metallicity evolution in the Universe. We tried two assumptions. In the first one, we assumed the stellar metallicity is equal to the mass-weighted average metallicity in each era (single metallicity evolution model) \citep{2017ApJ...840...39M}. In the other one, we considered the metallicity dispersion in each era as well as metallicity evolution \citep{2019MNRAS.488.5300C}. With metallicity dispersion in each formation era, the merger rate density calculated from our results was $70~{\rm yr}^{-1}{\rm Gpc}^{-3}$, which is twice as large as the case without metallicity dispersion. The merger rate density that we obtained is comparable to that estimated from the first and second runs of LIGO and Virgo \citep{2019ApJ...882L..24A}. We found that the effect of lower-metallicity clusters formed at low-$z$ was especially significant.

In addition, we investigated differential merger rate density for $M_1$ and $q$. We found that differential merger rate density calculated from our simulations generally agrees with that estimated from observed BBH mergers. However, BBH mergers with the primary mass larger than $30M_\odot$ in our simulations are several times less than the probability distribution estimated from the observed BBH mergers. Such BBHs may be formed in globular clusters and/or isolated field binaries formed at high-$z$.


\section*{Acknowledgements}

This work is supported by JSPS KAKENHI Grant Number 17H06360, 19H01933, and 19K03907 and The University of Tokyo Excellent Young Researcher Program. Numerical calculations reported in this paper were supported by Initiative on Promotion of Supercomputing for Young or Women Researchers, Supercomputing Division, Information Technology Center, The University of Tokyo. Finally, we thank the referees for giving constructive comments that helped us to improve this manuscript.


\bibliographystyle{mnras}
\bibliography{main} 

\begin{thebibliography}{}
\makeatletter
\relax
\def\mn@urlcharsother{\let\do\@makeother \do\$\do\&\do\#\do\^\do\_\do\%\do\~}
\def\mn@doi{\begingroup\mn@urlcharsother \@ifnextchar [ {\mn@doi@}
  {\mn@doi@[]}}
\def\mn@doi@[#1]#2{\def\@tempa{#1}\ifx\@tempa\@empty \href
  {http://dx.doi.org/#2} {doi:#2}\else \href {http://dx.doi.org/#2} {#1}\fi
  \endgroup}
\def\mn@eprint#1#2{\mn@eprint@#1:#2::\@nil}
\def\mn@eprint@arXiv#1{\href {http://arxiv.org/abs/#1} {{\tt arXiv:#1}}}
\def\mn@eprint@dblp#1{\href {http://dblp.uni-trier.de/rec/bibtex/#1.xml}
  {dblp:#1}}
\def\mn@eprint@#1:#2:#3:#4\@nil{\def\@tempa {#1}\def\@tempb {#2}\def\@tempc
  {#3}\ifx \@tempc \@empty \let \@tempc \@tempb \let \@tempb \@tempa \fi \ifx
  \@tempb \@empty \def\@tempb {arXiv}\fi \@ifundefined
  {mn@eprint@\@tempb}{\@tempb:\@tempc}{\expandafter \expandafter \csname
  mn@eprint@\@tempb\endcsname \expandafter{\@tempc}}}

\bibitem[\protect\citeauthoryear{{Aarseth}}{{Aarseth}}{1999}]{1999PASP..111.1333A}
{Aarseth} S.~J.,  1999, \mn@doi [\pasp] {10.1086/316455}, \href
  {http://adsabs.harvard.edu/abs/1999PASP..111.1333A} {111, 1333}

\bibitem[\protect\citeauthoryear{{Abbott} et~al.,}{{Abbott}
  et~al.}{2016a}]{2016PhRvL.116f1102A}
{Abbott} B.~P.,  et~al., 2016a, \mn@doi [Physical Review Letters]
  {10.1103/PhysRevLett.116.061102}, \href
  {http://adsabs.harvard.edu/abs/2016PhRvL.116f1102A} {116, 061102}

\bibitem[\protect\citeauthoryear{{Abbott} et~al.,}{{Abbott}
  et~al.}{2016b}]{2016PhRvL.116x1103A}
{Abbott} B.~P.,  et~al., 2016b, \mn@doi [Physical Review Letters]
  {10.1103/PhysRevLett.116.241103}, \href
  {http://adsabs.harvard.edu/abs/2016PhRvL.116x1103A} {116, 241103}

\bibitem[\protect\citeauthoryear{{Abbott} et~al.,}{{Abbott}
  et~al.}{2017a}]{2017PhRvL.118v1101A}
{Abbott} B.~P.,  et~al., 2017a, \mn@doi [Physical Review Letters]
  {10.1103/PhysRevLett.118.221101}, \href
  {http://adsabs.harvard.edu/abs/2017PhRvL.118v1101A} {118, 221101}

\bibitem[\protect\citeauthoryear{{Abbott} et~al.,}{{Abbott}
  et~al.}{2017b}]{2017PhRvL.119n1101A}
{Abbott} B.~P.,  et~al., 2017b, \mn@doi [Physical Review Letters]
  {10.1103/PhysRevLett.119.141101}, \href
  {http://adsabs.harvard.edu/abs/2017PhRvL.119n1101A} {119, 141101}

\bibitem[\protect\citeauthoryear{{Abbott} et~al.,}{{Abbott}
  et~al.}{2017c}]{2017ApJ...851L..35A}
{Abbott} B.~P.,  et~al., 2017c, \mn@doi [\apjl] {10.3847/2041-8213/aa9f0c},
  \href {http://adsabs.harvard.edu/abs/2017ApJ...851L..35A} {851, L35}

\bibitem[\protect\citeauthoryear{{Abbott} et~al.,}{{Abbott}
  et~al.}{2019a}]{2019PhRvX...9c1040A}
{Abbott} B.~P.,  et~al., 2019a, \mn@doi [Physical Review X]
  {10.1103/PhysRevX.9.031040}, \href
  {https://ui.adsabs.harvard.edu/abs/2019PhRvX...9c1040A} {9, 031040}

\bibitem[\protect\citeauthoryear{{Abbott} et~al.,}{{Abbott}
  et~al.}{2019b}]{2019ApJ...882L..24A}
{Abbott} B.~P.,  et~al., 2019b, \mn@doi [\apjl] {10.3847/2041-8213/ab3800},
  \href {https://ui.adsabs.harvard.edu/abs/2019ApJ...882L..24A} {882, L24}

\bibitem[\protect\citeauthoryear{{Antonini} \& {Rasio}}{{Antonini} \&
  {Rasio}}{2016}]{2016ApJ...831..187A}
{Antonini} F.,  {Rasio} F.~A.,  2016, \mn@doi [\apj]
  {10.3847/0004-637X/831/2/187}, \href
  {http://adsabs.harvard.edu/abs/2016ApJ...831..187A} {831, 187}

\bibitem[\protect\citeauthoryear{{Askar}, {Szkudlarek}, {Gondek-Rosi{\'n}ska},
  {Giersz}  \& {Bulik}}{{Askar} et~al.}{2017}]{2017MNRAS.464L..36A}
{Askar} A.,  {Szkudlarek} M.,  {Gondek-Rosi{\'n}ska} D.,  {Giersz} M.,
  {Bulik} T.,  2017, \mn@doi [\mnras] {10.1093/mnrasl/slw177}, \href
  {http://adsabs.harvard.edu/abs/2017MNRAS.464L..36A} {464, L36}

\bibitem[\protect\citeauthoryear{{Bae}, {Kim}  \& {Lee}}{{Bae}
  et~al.}{2014}]{2014MNRAS.440.2714B}
{Bae} Y.-B.,  {Kim} C.,   {Lee} H.~M.,  2014, \mn@doi [\mnras]
  {10.1093/mnras/stu381}, \href
  {http://adsabs.harvard.edu/abs/2014MNRAS.440.2714B} {440, 2714}

\bibitem[\protect\citeauthoryear{{Banerjee}}{{Banerjee}}{2017}]{2017MNRAS.467..524B}
{Banerjee} S.,  2017, \mn@doi [\mnras] {10.1093/mnras/stw3392}, \href
  {http://adsabs.harvard.edu/abs/2017MNRAS.467..524B} {467, 524}

\bibitem[\protect\citeauthoryear{{Banerjee}}{{Banerjee}}{2018a}]{2018MNRAS.473..909B}
{Banerjee} S.,  2018a, \mn@doi [\mnras] {10.1093/mnras/stx2347}, \href
  {http://adsabs.harvard.edu/abs/2018MNRAS.473..909B} {473, 909}

\bibitem[\protect\citeauthoryear{{Banerjee}}{{Banerjee}}{2018b}]{2018MNRAS.481.5123B}
{Banerjee} S.,  2018b, \mn@doi [\mnras] {10.1093/mnras/sty2608}, \href
  {http://adsabs.harvard.edu/abs/2018MNRAS.481.5123B} {481, 5123}

\bibitem[\protect\citeauthoryear{{Banerjee}, {Baumgardt}  \&
  {Kroupa}}{{Banerjee} et~al.}{2010}]{2010MNRAS.402..371B}
{Banerjee} S.,  {Baumgardt} H.,   {Kroupa} P.,  2010, \mn@doi [\mnras]
  {10.1111/j.1365-2966.2009.15880.x}, \href
  {http://adsabs.harvard.edu/abs/2010MNRAS.402..371B} {402, 371}

\bibitem[\protect\citeauthoryear{{Bavera} et~al.,}{{Bavera}
  et~al.}{2019}]{2019arXiv190612257B}
{Bavera} S.~S.,  et~al., 2019, arXiv e-prints, \href
  {https://ui.adsabs.harvard.edu/abs/2019arXiv190612257B} {p. arXiv:1906.12257}

\bibitem[\protect\citeauthoryear{{Belczynski}, {Bulik}, {Fryer}, {Ruiter},
  {Valsecchi}, {Vink}  \& {Hurley}}{{Belczynski}
  et~al.}{2010}]{Belczynski+2010}
{Belczynski} K.,  {Bulik} T.,  {Fryer} C.~L.,  {Ruiter} A.,  {Valsecchi} F.,
  {Vink} J.~S.,   {Hurley} J.~R.,  2010, \mn@doi [\apj]
  {10.1088/0004-637X/714/2/1217}, \href
  {http://adsabs.harvard.edu/abs/2010ApJ...714.1217B} {714, 1217}

\bibitem[\protect\citeauthoryear{{Belczynski}, {Holz}, {Bulik}  \&
  {O'Shaughnessy}}{{Belczynski} et~al.}{2016}]{2016Natur.534..512B}
{Belczynski} K.,  {Holz} D.~E.,  {Bulik} T.,   {O'Shaughnessy} R.,  2016,
  \mn@doi [\nat] {10.1038/nature18322}, \href
  {http://adsabs.harvard.edu/abs/2016Natur.534..512B} {534, 512}

\bibitem[\protect\citeauthoryear{{Bethe} \& {Brown}}{{Bethe} \&
  {Brown}}{1998}]{1998ApJ...506..780B}
{Bethe} H.~A.,  {Brown} G.~E.,  1998, \mn@doi [\apj] {10.1086/306265}, \href
  {http://adsabs.harvard.edu/abs/1998ApJ...506..780B} {506, 780}

\bibitem[\protect\citeauthoryear{{Bouffanais}, {Mapelli}, {Gerosa}, {Di Carlo},
  {Giacobbo}, {Berti}  \& {Baibhav}}{{Bouffanais}
  et~al.}{2019}]{2019ApJ...886...25B}
{Bouffanais} Y.,  {Mapelli} M.,  {Gerosa} D.,  {Di Carlo} U.~N.,  {Giacobbo}
  N.,  {Berti} E.,   {Baibhav} V.,  2019, \mn@doi [\apj]
  {10.3847/1538-4357/ab4a79}, \href
  {https://ui.adsabs.harvard.edu/abs/2019ApJ...886...25B} {886, 25}

\bibitem[\protect\citeauthoryear{{Chruslinska} \& {Nelemans}}{{Chruslinska} \&
  {Nelemans}}{2019}]{2019MNRAS.488.5300C}
{Chruslinska} M.,  {Nelemans} G.,  2019, \mn@doi [\mnras]
  {10.1093/mnras/stz2057}, \href
  {https://ui.adsabs.harvard.edu/abs/2019MNRAS.488.5300C} {488, 5300}

\bibitem[\protect\citeauthoryear{{Di Carlo}, {Giacobbo}, {Mapelli}, {Pasquato},
  {Spera}, {Wang}  \& {Haardt}}{{Di Carlo} et~al.}{2019}]{2019MNRAS.487.2947D}
{Di Carlo} U.~N.,  {Giacobbo} N.,  {Mapelli} M.,  {Pasquato} M.,  {Spera} M.,
  {Wang} L.,   {Haardt} F.,  2019, \mn@doi [\mnras] {10.1093/mnras/stz1453},
  \href {https://ui.adsabs.harvard.edu/abs/2019MNRAS.487.2947D} {487, 2947}

\bibitem[\protect\citeauthoryear{{Dominik}, {Belczynski}, {Fryer}, {Holz},
  {Berti}, {Bulik}, {Mand el}  \& {O'Shaughnessy}}{{Dominik}
  et~al.}{2012}]{2012ApJ...759...52D}
{Dominik} M.,  {Belczynski} K.,  {Fryer} C.,  {Holz} D.~E.,  {Berti} E.,
  {Bulik} T.,  {Mand el} I.,   {O'Shaughnessy} R.,  2012, \mn@doi [\apj]
  {10.1088/0004-637X/759/1/52}, \href
  {https://ui.adsabs.harvard.edu/abs/2012ApJ...759...52D} {759, 52}

\bibitem[\protect\citeauthoryear{{Dominik} et~al.,}{{Dominik}
  et~al.}{2015}]{2015ApJ...806..263D}
{Dominik} M.,  et~al., 2015, \mn@doi [\apj] {10.1088/0004-637X/806/2/263},
  \href {https://ui.adsabs.harvard.edu/abs/2015ApJ...806..263D} {806, 263}

\bibitem[\protect\citeauthoryear{{Downing}, {Benacquista}, {Giersz}  \&
  {Spurzem}}{{Downing} et~al.}{2010}]{2010MNRAS.407.1946D}
{Downing} J.~M.~B.,  {Benacquista} M.~J.,  {Giersz} M.,   {Spurzem} R.,  2010,
  \mn@doi [\mnras] {10.1111/j.1365-2966.2010.17040.x}, \href
  {http://adsabs.harvard.edu/abs/2010MNRAS.407.1946D} {407, 1946}

\bibitem[\protect\citeauthoryear{{Downing}, {Benacquista}, {Giersz}  \&
  {Spurzem}}{{Downing} et~al.}{2011}]{2011MNRAS.416..133D}
{Downing} J.~M.~B.,  {Benacquista} M.~J.,  {Giersz} M.,   {Spurzem} R.,  2011,
  \mn@doi [\mnras] {10.1111/j.1365-2966.2011.19023.x}, \href
  {http://adsabs.harvard.edu/abs/2011MNRAS.416..133D} {416, 133}

\bibitem[\protect\citeauthoryear{{Fujii} \& {Portegies Zwart}}{{Fujii} \&
  {Portegies Zwart}}{2016}]{2016ApJ...817....4F}
{Fujii} M.~S.,  {Portegies Zwart} S.,  2016, \mn@doi [\apj]
  {10.3847/0004-637X/817/1/4}, \href
  {http://adsabs.harvard.edu/abs/2016ApJ...817....4F} {817, 4}

\bibitem[\protect\citeauthoryear{{Fujii}, {Tanikawa}  \& {Makino}}{{Fujii}
  et~al.}{2017}]{2017PASJ...69...94F}
{Fujii} M.~S.,  {Tanikawa} A.,   {Makino} J.,  2017, \mn@doi [\pasj]
  {10.1093/pasj/psx108}, \href
  {http://adsabs.harvard.edu/abs/2017PASJ...69...94F} {69, 94}

\bibitem[\protect\citeauthoryear{{Gaia Collaboration} et~al.,}{{Gaia
  Collaboration} et~al.}{2016}]{GaiaColl2016}
{Gaia Collaboration} et~al., 2016, \mn@doi [\aap]
  {10.1051/0004-6361/201629272}, \href
  {https://ui.adsabs.harvard.edu/abs/2016A&A...595A...1G} {595, A1}

\bibitem[\protect\citeauthoryear{{Giacobbo}, {Mapelli}  \& {Spera}}{{Giacobbo}
  et~al.}{2018}]{2018MNRAS.474.2959G}
{Giacobbo} N.,  {Mapelli} M.,   {Spera} M.,  2018, \mn@doi [\mnras]
  {10.1093/mnras/stx2933}, \href
  {https://ui.adsabs.harvard.edu/abs/2018MNRAS.474.2959G} {474, 2959}

\bibitem[\protect\citeauthoryear{{Goswami}, {Kiel}  \& {Rasio}}{{Goswami}
  et~al.}{2014}]{2014ApJ...781...81G}
{Goswami} S.,  {Kiel} P.,   {Rasio} F.~A.,  2014, \mn@doi [\apj]
  {10.1088/0004-637X/781/2/81}, \href
  {http://adsabs.harvard.edu/abs/2014ApJ...781...81G} {781, 81}

\bibitem[\protect\citeauthoryear{{G{\"u}rkan}, {Freitag}  \&
  {Rasio}}{{G{\"u}rkan} et~al.}{2004}]{2004ApJ...604..632G}
{G{\"u}rkan} M.~A.,  {Freitag} M.,   {Rasio} F.~A.,  2004, \mn@doi [\apj]
  {10.1086/381968}, \href {http://adsabs.harvard.edu/abs/2004ApJ...604..632G}
  {604, 632}

\bibitem[\protect\citeauthoryear{{Hong}, {Vesperini}, {Askar}, {Giersz},
  {Szkudlarek}  \& {Bulik}}{{Hong} et~al.}{2018}]{2018MNRAS.480.5645H}
{Hong} J.,  {Vesperini} E.,  {Askar} A.,  {Giersz} M.,  {Szkudlarek} M.,
  {Bulik} T.,  2018, \mn@doi [\mnras] {10.1093/mnras/sty2211}, \href
  {http://adsabs.harvard.edu/abs/2018MNRAS.480.5645H} {480, 5645}

\bibitem[\protect\citeauthoryear{{Hurley}, {Pols}  \& {Tout}}{{Hurley}
  et~al.}{2000}]{Hurley+2000}
{Hurley} J.~R.,  {Pols} O.~R.,   {Tout} C.~A.,  2000, \mn@doi [\mnras]
  {10.1046/j.1365-8711.2000.03426.x}, \href
  {http://adsabs.harvard.edu/abs/2000MNRAS.315..543H} {315, 543}

\bibitem[\protect\citeauthoryear{{Kinugawa}, {Inayoshi}, {Hotokezaka},
  {Nakauchi}  \& {Nakamura}}{{Kinugawa} et~al.}{2014}]{2014MNRAS.442.2963K}
{Kinugawa} T.,  {Inayoshi} K.,  {Hotokezaka} K.,  {Nakauchi} D.,   {Nakamura}
  T.,  2014, \mn@doi [\mnras] {10.1093/mnras/stu1022}, \href
  {http://adsabs.harvard.edu/abs/2014MNRAS.442.2963K} {442, 2963}

\bibitem[\protect\citeauthoryear{{Kroupa}}{{Kroupa}}{2001}]{Kroupa2001}
{Kroupa} P.,  2001, \mn@doi [\mnras] {10.1046/j.1365-8711.2001.04022.x}, \href
  {http://adsabs.harvard.edu/abs/2001MNRAS.322..231K} {322, 231}

\bibitem[\protect\citeauthoryear{{Kumamoto}, {Fujii}  \& {Tanikawa}}{{Kumamoto}
  et~al.}{2019}]{2019MNRAS.486.3942K}
{Kumamoto} J.,  {Fujii} M.~S.,   {Tanikawa} A.,  2019, \mn@doi [\mnras]
  {10.1093/mnras/stz1068}, \href
  {https://ui.adsabs.harvard.edu/abs/2019MNRAS.486.3942K} {486, 3942}

\bibitem[\protect\citeauthoryear{{Kustaanheimo} \& {Stiefel}}{{Kustaanheimo} \&
  {Stiefel}}{1965}]{KustaanheimoStiefel1965}
{Kustaanheimo} P.,  {Stiefel} E.,  1965, \mn@doi [Journal f{\"u}r die reine und
  angewandte Mathematik] {10.1515/crll.1965.218.204}, 218, 204

\bibitem[\protect\citeauthoryear{{Madau} \& {Fragos}}{{Madau} \&
  {Fragos}}{2017}]{2017ApJ...840...39M}
{Madau} P.,  {Fragos} T.,  2017, \mn@doi [\apj] {10.3847/1538-4357/aa6af9},
  \href {https://ui.adsabs.harvard.edu/abs/2017ApJ...840...39M} {840, 39}

\bibitem[\protect\citeauthoryear{{Makino} \& {Aarseth}}{{Makino} \&
  {Aarseth}}{1992}]{1992PASJ...44..141M}
{Makino} J.,  {Aarseth} S.~J.,  1992, \pasj, \href
  {http://adsabs.harvard.edu/abs/1992PASJ...44..141M} {44, 141}

\bibitem[\protect\citeauthoryear{{Mapelli}}{{Mapelli}}{2016}]{2016MNRAS.459.3432M}
{Mapelli} M.,  2016, \mn@doi [\mnras] {10.1093/mnras/stw869}, \href
  {http://adsabs.harvard.edu/abs/2016MNRAS.459.3432M} {459, 3432}

\bibitem[\protect\citeauthoryear{{Mikkola} \& {Aarseth}}{{Mikkola} \&
  {Aarseth}}{1993}]{1993CeMDA..57..439M}
{Mikkola} S.,  {Aarseth} S.~J.,  1993, \mn@doi [Celestial Mechanics and
  Dynamical Astronomy] {10.1007/BF00695714}, \href
  {http://adsabs.harvard.edu/abs/1993CeMDA..57..439M} {57, 439}

\bibitem[\protect\citeauthoryear{{Morscher}, {Umbreit}, {Farr}  \&
  {Rasio}}{{Morscher} et~al.}{2013}]{2013ApJ...763L..15M}
{Morscher} M.,  {Umbreit} S.,  {Farr} W.~M.,   {Rasio} F.~A.,  2013, \mn@doi
  [\apjl] {10.1088/2041-8205/763/1/L15}, \href
  {http://adsabs.harvard.edu/abs/2013ApJ...763L..15M} {763, L15}

\bibitem[\protect\citeauthoryear{{O'Leary}, {Rasio}, {Fregeau}, {Ivanova}  \&
  {O'Shaughnessy}}{{O'Leary} et~al.}{2006}]{2006ApJ...637..937O}
{O'Leary} R.~M.,  {Rasio} F.~A.,  {Fregeau} J.~M.,  {Ivanova} N.,
  {O'Shaughnessy} R.,  2006, \mn@doi [\apj] {10.1086/498446}, \href
  {http://adsabs.harvard.edu/abs/2006ApJ...637..937O} {637, 937}

\bibitem[\protect\citeauthoryear{{O'Leary}, {Kocsis}  \& {Loeb}}{{O'Leary}
  et~al.}{2009}]{2009MNRAS.395.2127O}
{O'Leary} R.~M.,  {Kocsis} B.,   {Loeb} A.,  2009, \mn@doi [\mnras]
  {10.1111/j.1365-2966.2009.14653.x}, \href
  {http://adsabs.harvard.edu/abs/2009MNRAS.395.2127O} {395, 2127}

\bibitem[\protect\citeauthoryear{{Park}, {Kim}, {Lee}, {Bae}  \&
  {Belczynski}}{{Park} et~al.}{2017}]{2017MNRAS.469.4665P}
{Park} D.,  {Kim} C.,  {Lee} H.~M.,  {Bae} Y.-B.,   {Belczynski} K.,  2017,
  \mn@doi [\mnras] {10.1093/mnras/stx1015}, \href
  {http://adsabs.harvard.edu/abs/2017MNRAS.469.4665P} {469, 4665}

\bibitem[\protect\citeauthoryear{{Peters} \& {Mathews}}{{Peters} \&
  {Mathews}}{1963}]{1963PhRv..131..435P}
{Peters} P.~C.,  {Mathews} J.,  1963, \mn@doi [Physical Review]
  {10.1103/PhysRev.131.435}, \href
  {http://adsabs.harvard.edu/abs/1963PhRv..131..435P} {131, 435}

\bibitem[\protect\citeauthoryear{{Plummer}}{{Plummer}}{1911}]{1911MNRAS..71..460P}
{Plummer} H.~C.,  1911, \mn@doi [\mnras] {10.1093/mnras/71.5.460}, \href
  {http://adsabs.harvard.edu/abs/1911MNRAS..71..460P} {71, 460}

\bibitem[\protect\citeauthoryear{{Portegies Zwart} \& {McMillan}}{{Portegies
  Zwart} \& {McMillan}}{2000}]{2000ApJ...528L..17P}
{Portegies Zwart} S.~F.,  {McMillan} S.~L.~W.,  2000, \mn@doi [\apjl]
  {10.1086/312422}, \href {http://adsabs.harvard.edu/abs/2000ApJ...528L..17P}
  {528, L17}

\bibitem[\protect\citeauthoryear{{Portegies Zwart}, {McMillan}  \&
  {Gieles}}{{Portegies Zwart} et~al.}{2010}]{2010ARA&A..48..431P}
{Portegies Zwart} S.~F.,  {McMillan} S.~L.~W.,   {Gieles} M.,  2010, \mn@doi
  [\araa] {10.1146/annurev-astro-081309-130834}, \href
  {http://adsabs.harvard.edu/abs/2010ARA%26A..48..431P} {48, 431}

\bibitem[\protect\citeauthoryear{{Rastello}, {Amaro-Seoane}, {Arca-Sedda},
  {Capuzzo-Dolcetta}, {Fragione}  \& {Tosta e Melo}}{{Rastello}
  et~al.}{2019}]{2019MNRAS.483.1233R}
{Rastello} S.,  {Amaro-Seoane} P.,  {Arca-Sedda} M.,  {Capuzzo-Dolcetta} R.,
  {Fragione} G.,   {Tosta e Melo} I.,  2019, \mn@doi [\mnras]
  {10.1093/mnras/sty3193}, \href
  {http://adsabs.harvard.edu/abs/2019MNRAS.483.1233R} {483, 1233}

\bibitem[\protect\citeauthoryear{{Rodriguez}, {Morscher}, {Pattabiraman},
  {Chatterjee}, {Haster}  \& {Rasio}}{{Rodriguez}
  et~al.}{2015}]{2015PhRvL.115e1101R}
{Rodriguez} C.~L.,  {Morscher} M.,  {Pattabiraman} B.,  {Chatterjee} S.,
  {Haster} C.-J.,   {Rasio} F.~A.,  2015, \mn@doi [Physical Review Letters]
  {10.1103/PhysRevLett.115.051101}, \href
  {http://adsabs.harvard.edu/abs/2015PhRvL.115e1101R} {115, 051101}

\bibitem[\protect\citeauthoryear{{Rodriguez}, {Chatterjee}  \&
  {Rasio}}{{Rodriguez} et~al.}{2016}]{2016PhRvD..93h4029R}
{Rodriguez} C.~L.,  {Chatterjee} S.,   {Rasio} F.~A.,  2016, \mn@doi [\prd]
  {10.1103/PhysRevD.93.084029}, \href
  {http://adsabs.harvard.edu/abs/2016PhRvD..93h4029R} {93, 084029}

\bibitem[\protect\citeauthoryear{{Sadowski}, {Belczynski}, {Bulik}, {Ivanova},
  {Rasio}  \& {O'Shaughnessy}}{{Sadowski} et~al.}{2008}]{2008ApJ...676.1162S}
{Sadowski} A.,  {Belczynski} K.,  {Bulik} T.,  {Ivanova} N.,  {Rasio} F.~A.,
  {O'Shaughnessy} R.,  2008, \mn@doi [\apj] {10.1086/528932}, \href
  {http://adsabs.harvard.edu/abs/2008ApJ...676.1162S} {676, 1162}

\bibitem[\protect\citeauthoryear{{Tanikawa}}{{Tanikawa}}{2013}]{2013MNRAS.435.1358T}
{Tanikawa} A.,  2013, \mn@doi [\mnras] {10.1093/mnras/stt1380}, \href
  {http://adsabs.harvard.edu/abs/2013MNRAS.435.1358T} {435, 1358}

\bibitem[\protect\citeauthoryear{{Tout}, {Aarseth}, {Pols}  \&
  {Eggleton}}{{Tout} et~al.}{1997}]{1997MNRAS.291..732T}
{Tout} C.~A.,  {Aarseth} S.~J.,  {Pols} O.~R.,   {Eggleton} P.~P.,  1997,
  \mn@doi [\mnras] {10.1093/mnras/291.4.732}, \href
  {http://adsabs.harvard.edu/abs/1997MNRAS.291..732T} {291, 732}

\bibitem[\protect\citeauthoryear{{Tutukov}, {Yungelson}  \&
  {Klayman}}{{Tutukov} et~al.}{1973}]{1973NInfo..27....3T}
{Tutukov} A.,  {Yungelson} L.,   {Klayman} A.,  1973, Nauchnye Informatsii,
  \href {http://adsabs.harvard.edu/abs/1973NInfo..27....3T} {27, 3}

\bibitem[\protect\citeauthoryear{{Wang}, {Spurzem}, {Aarseth}, {Nitadori},
  {Berczik}, {Kouwenhoven}  \& {Naab}}{{Wang} et~al.}{2015}]{Wang+2015}
{Wang} L.,  {Spurzem} R.,  {Aarseth} S.,  {Nitadori} K.,  {Berczik} P.,
  {Kouwenhoven} M.~B.~N.,   {Naab} T.,  2015, \mn@doi [\mnras]
  {10.1093/mnras/stv817}, \href
  {http://adsabs.harvard.edu/abs/2015MNRAS.450.4070W} {450, 4070}

\bibitem[\protect\citeauthoryear{{Zevin}, {Samsing}, {Rodriguez}, {Haster}  \&
  {Ramirez-Ruiz}}{{Zevin} et~al.}{2019}]{2019ApJ...871...91Z}
{Zevin} M.,  {Samsing} J.,  {Rodriguez} C.,  {Haster} C.-J.,   {Ramirez-Ruiz}
  E.,  2019, \mn@doi [\apj] {10.3847/1538-4357/aaf6ec}, \href
  {https://ui.adsabs.harvard.edu/abs/2019ApJ...871...91Z} {871, 91}

\bibitem[\protect\citeauthoryear{{Ziosi}, {Mapelli}, {Branchesi}  \&
  {Tormen}}{{Ziosi} et~al.}{2014}]{2014MNRAS.441.3703Z}
{Ziosi} B.~M.,  {Mapelli} M.,  {Branchesi} M.,   {Tormen} G.,  2014, \mn@doi
  [\mnras] {10.1093/mnras/stu824}, \href
  {http://adsabs.harvard.edu/abs/2014MNRAS.441.3703Z} {441, 3703}

\makeatother
\end{thebibliography}


%
%


\bsp	
\label{lastpage}
\end{document}